\documentclass[journal]{IEEEtran}
\usepackage{amsmath,amsfonts,graphicx,tabularx,color}
\usepackage{amssymb,textcomp,mathtools,amsthm,gensymb}
\usepackage{bm,upgreek,algorithm,hyperref}
\usepackage{multirow,booktabs,hhline,array}
\usepackage{cite,url,makecell,setspace,listings,verbatim}
\RequirePackage{doi}

\definecolor{codegreen}{rgb}{0,0.6,0}
\definecolor{codegray}{rgb}{0.5,0.5,0.5}
\definecolor{codepurple}{rgb}{0.58,0,0.82}
\definecolor{backcolour}{rgb}{0.95,0.95,0.92}

\renewcommand{\vec}[1]{\bm{\mathrm{#1}}}

\theoremstyle{definition}

\theoremstyle{plain}

\theoremstyle{remark}

\title{ReZero: Region-customizable Sound Extraction}

\author{Rongzhi~Gu, Yi~Luo}

\begin{document}
\maketitle

\begin{abstract}
We introduce region-customizable sound extraction (ReZero), a general and flexible framework for the multi-channel region-wise sound extraction (R-SE) task. R-SE task aims at extracting all active target sounds (e.g., human speech) within a specific, user-defined spatial region, which is different from conventional and existing tasks where a blind separation or a fixed, predefined spatial region are typically assumed. The spatial region can be defined as an angular window, a sphere, a cone, or other geometric patterns. Being a solution to the R-SE task, the proposed ReZero framework includes (1) definitions of different types of spatial regions, (2) methods for region feature extraction and aggregation, and (3) a multi-channel extension of the band-split RNN (BSRNN) model specified for the R-SE task. We design experiments for different microphone array geometries, different types of spatial regions, and comprehensive ablation studies on different system configurations. Experimental results on both simulated and real-recorded data demonstrate the effectiveness of ReZero. Demos are available at \url{https://innerselfm.github.io/rezero/}.

\end{abstract}


\section{Introduction}
\label{sec:intro}
\begin{figure}[!t]
  \centering
  \includegraphics[width=\linewidth]{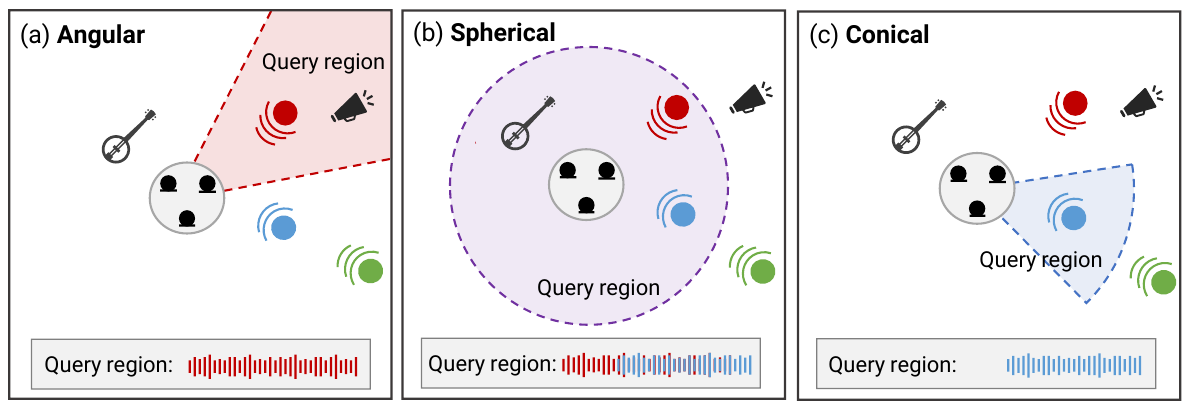}
\caption{Three typical cases of R-SE: (a) angular region; (b) spherical region; (c) conical region. The angle window of (a), radius of (b), and both angle window and radius of (c) can be dynamically assigned per needed. }
\label{fig:rse_app}
\end{figure}

Region-wise sound extraction (R-SE) has gained increased interest in recent years with a wide range of applications in selective hearing, offline conference, hearing aids, and audio augmented reality \cite{jenrungrot2020cone,xu2022learning,patterson2022distance,khandelwal2020two,nair2019audiovisual,anton2023dsenet}. Unlike conventional multi-channel source separation systems that aim at either blindly separating all active sources or extracting sounds coming from a certain direction or predefined region, R-SE attempts to extract active sources within a \textit{specific}, \textit{user-defined} spatial region, as shown in figure~\ref{fig:rse_app}. In figure~\ref{fig:rse_app} (a), the query region is angular when only the target sounds (e.g., human speech) within the angle window or direction range are desired. This can be useful when the target sources are located in a pre-arranged region and have a certain direction difference from other competing or interfering sources. Except for the angular region, the target region can also be a sphere that extracts sounds within a certain distance threshold, as illustrated in figure~\ref{fig:rse_app} (b). This scenario is suitable for removing distant speech or performing close-speaker extraction. For more fine-grained spatial regions, figure~\ref{fig:rse_app} (c) defines a conical region that considers both the direction range and the distance threshold.

One advantage of R-SE is that it relaxes the requirement for accurate target-source-related information to perform source extraction. Conventional source extraction methods rely on either a speaker enrollment or embedding for personalized speech extraction (P-SE) \cite{vzmolikova2019speakerbeam,wang2018voicefilter,emre22pdns} or a precise direction-of-arrival (DOA) or location for direction-aware speech extraction (D-SE) \cite{gu2019neural,chen2018efficient,gu20213d}. However, speaker enrollment or embedding may not be able to accurately match the characteristics of the target speaker in all recording conditions, and the accurate location information for the target sources might be hard to acquire. R-SE only needs a coarse region query, alleviating the requirement for such auxiliary information. 


In this paper, we propose a general and flexible framework for the R-SE task, which we refer to as \textit{\textbf{Re}}gion-customi\textit{\textbf{Z}}able sound \textit{\textbf{e}}xt\textit{\textbf{r}}acti\textit{\textbf{o}}n (\textit{\textbf{ReZero}}). ReZero aims at extracting all target sources, which we define as human speech, within a user-defined query region, by properly calculating region features and region descriptors. We define different region features for angle window and distance threshold, and introduce a modified multi-channel band-split RNN (BSRNN) \cite{luo2023music} network architecture, whose single-channel counterpart has been proven effective in music source separation and speech enhancement tasks \cite{yu2022high, luo2023music, yu2023efficient, yu2023tspeech}, to effectively make use of these region features. We design comprehensive experiments on different microphone array configurations, query region types, feature extraction methods, model sizes and complexities and competing systems, and the experiment results on both synthetic and real-recorded data show that ReZero is consistently better than other existing source separation and extraction methods.

The rest of the paper is organized as follows. Section~\ref{sec:related} briefly reviews the existing works on direction-based, distance-based and fixed-zone-based speech extraction. Section~\ref{sec:problem} formulates the problem of R-SE. Section~\ref{sec:rezero} introduces the components in our proposed ReZero framework, which include the feature extraction module, region feature sampling and aggregation module, and the neural network architecture designed for the R-SE task. Section~\ref{sec:config} describes the experiment configurations in detail. Section~\ref{sec:result} presents and analyzes the experiment results. Section~\ref{sec:conclusion} concludes the paper.

\section{Related works}
\label{sec:related}
We briefly review the existing related works in three aspects: 1) direction-based speech extraction (D-SE), which is a special case for angular-region-based sound extraction; 2) distance-based speech extraction, which is a special case for sphere-based sound extraction when the radius is fixed; 3) fixed-zone-based speech extraction, when the region shape and position are fixed. 

\subsection{Direction-based speech extraction}
The problem formulation of direction-based sound extraction is closely related to that of spatial filtering \cite{van1988beamforming} (e.g., beamforming) which aims to enhance the signal from a specific direction. While the most straightforward approach is to apply fixed or adaptive beamforming towards the target direction \cite{xiao18ICMV,khandelwal2020two}, there are three main limitations. First, although the beamformer formulations can be redesigned to adapt to different steering directions and mainlobe widths, their performance might be degraded when the target or interference sources are close to each other. Second, since the number of spatial nulls is constrained by the number of microphones \cite{levin2015average}, the ability for such beamformers to eliminate directional interference is thus limited. Third, such beamforming algorithms cannot fully cancel out isotropic or babble noise in the target direction. As an improvement to standard beamforming methods, neural networks equipped with such prior knowledge have been developed to conquer these issues in recent years \cite{chen2018efficient,chen2018multi,gu2019neural,tesch2022insights}. In such systems, the target direction is assumed to be available in advance or estimated with visual clues or audio localization techniques. The direction is then encoded into steered beams \cite{li2019direction,jenrungrot2020cone}, direction features \cite{chen2018multi, gu2019neural,gu2023towards} or used to initialize hidden states of RNNs \cite{tesch2021nonlinear,tesch2022insights}. Such methods have demonstrated to exhibit better performances compared to blind source separation (BSS) models. 

A special framework for D-SE is cone of silence (CoS) \cite{jenrungrot2020cone}, a Demucs-based \cite{defossez2019demucs} neural network in the waveform domain that iteratively separates sources within a gradually-narrowed-down angle window, given the center angle $\theta$ and pre-set window sizes $\{w_i\}_{i=1}^K$. The center angle is encoded into an enhanced waveform using delay-and-sum (DAS) beamforming, and the window size is embedded as a global conditioning variable to all the encoder and decoder blocks in the Demucs model. However, the model only considered 1D angular case where all the speech and noise were assumed to be on the same plane with the microphone array, and the iterative separation process makes it hard to balance system delay and complexity in streaming applications.



\subsection{Distance-based speech extraction}
Speakers located at different distances towards the microphones may have different energy or reverberation levels, which make distance-based speech extraction possible when such features can be properly designed and utilized. Recently proposed distance-based speech extraction approaches \cite{patterson2022distance,lin2023focus} enhance the near-field speech from monaural mixture signal within a \emph{pre-set} and \emph{fixed} distance threshold, e.g., 1.5 meters, to distinguish between ``near'' and ``far'' speakers. It was stated that the network implicitly learned to estimate the direct-to-reverberation ratio (DRR) of each speech and used such cues to separate the signals. However, existing models can only handle a fixed distance threshold rather than a user-defined distance query, and changing the distance threshold may result in retraining the entire model. Also, experiments were only conducted on noise-free simulated mixtures, which may cause mismatch to real-world scenarios.



\subsection{Fixed-zone-based speech extraction}
Speech extraction in fixed spatial zones or regions is naturally suitable for applications where potential speakers are located in pre-known regions, such as mobile phones where the front side is where speakers speak towards \cite{khandelwal2020two,anton2023dsenet}, smart glasses where the location of mouth is relatively easy to acquire \cite{markovic2022implicit}, and in-car scenarios where each seat can be treated as a fixed region \cite{wechsler2023multi,gu20213d,kothapally2023deep}. Such scenarios do no require region features as the locations of the regions are known and fixed, and one can train models to directly estimate target sources in each region. The proposed ReZero framework attempts to solve the  problem to allow the model to accept a customizable region query.



\section{Region-wise sound extraction}
\label{sec:problem}
We first describe the problem definition of the general R-SE task. The mixture signal received by a microphone array can be represented as:
\begin{equation}
    \vec{y}^m = \sum_{c=0}^{C-1} \vec{x}_c^m+ \vec{n}^m
\end{equation}
where $\vec{y}^m \in \mathbb{R}^{T}, m=1, \ldots, M$ denotes the mixture signal at the $m$-th channel, $T$ denotes the signal length, $C$ denotes the total number of the target sources, $\vec{n}$ denotes the sum of all point and isotropic noise signals, and $\vec{x}_c^m = \vec{x}_c^{m,d} + \vec{x}_c^{m,r}, \vec{x}_c^m \in \mathbb{R}^{T}$ denotes the $c$-th multi-channel reverberant target signal that can be split into the direct path and the early reflection component $\vec{x}_c^{m,d}$ and the late reverberation component $\vec{x}_c^{m,r}$. Each target signal $\vec{x}_c^m$ is associated with a precise location defined in polar coordinate $\{\theta_c,\phi_c, d_c\}$, respectively representing its azimuth and elevation with respect to a pre-defined coordinate system and distance with respect to the center of the array. 

In this paper, we focus on the task of simultaneously extracting the \emph{direct sound and early reflections} of all \emph{speech signals} and removing all noise signals within a query region defined by azimuth, elevation and distance ranges $\vec{r}=\{[\theta_l,\theta_h],[\phi_l,\phi_h], [d_l,d_h]\}$. The expected output of the system is then defined as:
\begin{equation}
    \label{eq:target}
    \begin{split}
        \vec{z}^{\text{ref}} &= \sum_{q=0}^{Q-1} \vec{x}_q^{\text{ref},d} \\
        \text{s.t. } \theta_l \leq \theta_q \leq \theta_h, &\phi_l \leq \phi_q \leq \phi_h, d_l\leq d_q \leq d_h
    \end{split}
\end{equation}
where $Q \in [0, C]$ is the number of speech signals within the query region, and $\vec{z}^{\text{ref}} \in \mathbb{R}^{T}$ denotes the target signal at a selected reference channel. This task formulation jointly performs speech extraction and dereverberation in a noisy environment, which matches a most common case in daily communication.

We consider three main region types depicted in figure \ref{fig:rse_app}: angle window (direction-only), sphere (distance-only), and cone (joint direction and distance). The angular region defines the case where $\vec{r}=\{[\theta_l,\theta_h],[\phi_l,\phi_h], [0,\infty]\}$, the spherical region defines the case where $\vec{r}=\{[-180\degree,180\degree],[-90\degree,90\degree], [0,d_h]\}$, and the conical region defines the case where $\vec{r}=\{[\theta_l,\theta_h],[\phi_l,\phi_h], [0,d_h]\}$.

\section{ReZero: a general framework for R-SE}
\label{sec:rezero}

\subsection{Pipeline overview}
\label{subsec:overview}

\begin{figure}[!ht]
  \centering
  \includegraphics[width=\linewidth]{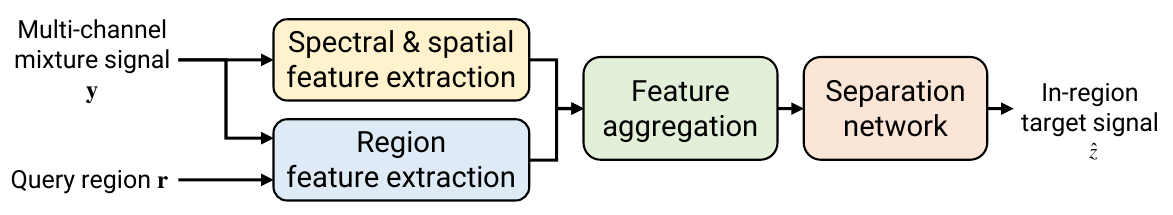}
\caption{The overview of the proposed ReZero framework.}
\label{fig:rezero}
\end{figure}

Figure~\ref{fig:rezero} provides the flowchart for the proposed region-customizable sound extraction (ReZero) framework. The input to the system includes the mixture signal $\{\vec{y}^m\}_{m=1}^M$ and the query region $\vec{r}$, where $\vec{r}$ is converted into a set of region features. The region features will then be aggregated to form a region descriptor and sent to an extraction network together with the spectral and spatial features calculated from the mixture signals. The extraction network then estimates the in-region target signal $\vec{z}^{\text{ref}}$ given all the features.

\subsection{Spatial and spectral feature extraction}
\label{subsec:ss_feature}

In our proposed system we operate in the time-frequency (T-F) domain. Following recent works on multi-channel speech separation \cite{wang2019combining,wang2018multi,gu2019neural,chen2018efficient, gu2023towards}, we take the complex spectrogram for the mixture signals as the spectral feature, and the interaural phase difference (IPD) and interaural level difference (ILD) as the spatial features:
\begin{equation} 
\begin{split}
    \text{IPD}^{(p)}(t,f) &= \angle Y^{p_1}(t,f)-\angle Y^{p_2}(t,f) \\
    \text{ILD}^{(p)}(t,f) &= 20\log{\frac{|Y^{p_1}(t,f)|}{|Y^{p_2}(t,f)|} } \\
\end{split}
\label{eq:tf_spatial}
\end{equation}
where $p=(p_1, p_2)$ denotes the microphone pair index, and $Y(t,f) \in \mathbb{C}^M$ denotes the complex-valued T-F bin at time $t$ and frequency $f$ for signal $\vec{y}$.


\subsection{Region feature extraction}
\label{subsec:region_feature}

We define direction and distance features in different ways. For direction feature, we follow previous studies on direction-based speech separation and fixed-zone-based speech extraction \cite{gu2019neural, gu20213d} where the similarity between IPD and target phase difference (TPD) within the query angle window is used as the direction feature. For distance feature, we use a distance embedding generator (DEG) to generate learnable distance embeddings.

\subsubsection{Direction feature}
Given an azimuth $\theta$, an elevation $\phi$ and a microphone pair index $p$, we extract the feature at this specific direction $V(\theta,\phi,t,f) \in \mathbb{R}$ by \cite{chen2018multi,gu2019neural}:
\begin{equation}
    \begin{split}        
        V(\theta,\phi,t,f) &= {\sum}_p\left <\mathbf{e}^{\text{IPD}^{(p)}(t,f)}, \mathbf{e}^{\text{TPD}^{(p)}(\theta,\phi,f)} \right >     \\
        \text{TPD}^{(p)}(\theta,\phi,f) &= 2\pi f\tau^{(p)}(\theta,\phi) \\
        \tau^{(p)}(\theta,\phi) &= d^{(p)}(\theta,\phi) f_s / v \\
        d^{(p)}(\theta,\phi) &= \varDelta^{(p)}\cos{\theta}\cos{\phi}
    \end{split}
    \label{eq:azm_DF}
\end{equation}
where vector $\mathbf{e}^{(\cdot)}=\begin{bmatrix} \cos(\cdot) \\ \sin(\cdot) \end{bmatrix}$ calculates the cosine and sine of the angles and stack them to form a 2-D vector, $\left<\cdot\right>$ denotes inner product, $\tau^{(p)}(\theta,\phi)$ corresponds to the theoretical delay that a unit impulse may experience between the $p$-th microphone pair, $\varDelta^{(p)}$ and $d^{(p)}(\theta,\phi)$ are the spacing and time difference of arrival (TDOA) of the $p$-th microphone pair \cite{gustafsson2003positioning,gu20213d}, respectively,  $f_s$ denotes the sampling rate, and $v$ denotes the sound velocity. 
$V(\theta,\phi,t,f)$ measures the similarity between the theoretical and observed phase differences at a certain T-F bin and direction \cite{gu2023towards}. A higher similarity score indicates that the observed signal has a higher chance to have sources coming from this selected direction $\{\theta,\phi\}$.


\subsubsection{Distance feature}
\label{subsubsec:sphere_feature}
\begin{figure}[!ht]
  \centering
  \includegraphics[width=7cm]{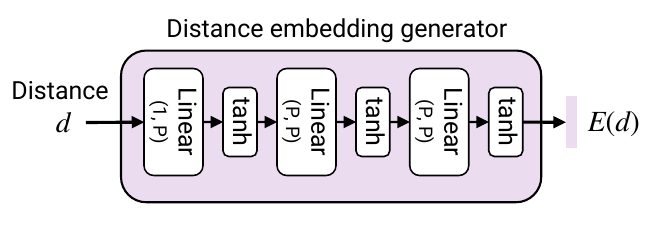}
\caption{The illustration of the distance embedding generator (DEG), which maps a distance $d$ to a distance embedding $E(d)$.}
\label{fig:sphere}
\end{figure}

We use a distance embedding generator (DEG), in the form of a simple multi-layer perceptron (MLP), to generate distance feature given a distance $d$. DEG takes the scalar $d$ as input, which is similar to recent works on Hypernetworks \cite{ha2017hypernetworks, chen2018neural}, and generates an embedding $E(d) \in \mathbb{R}^{P}$ that represents this particular distance threshold. Unlike direction features which are purely defined on signal statistics, the DEG network is jointly optimized with the rest of the system to allow end-to-end optimization.


\subsection{Region feature sampling}
\label{subsubsec:region_sample}

\begin{figure}[!ht]
  \centering
  \includegraphics[width=\linewidth]{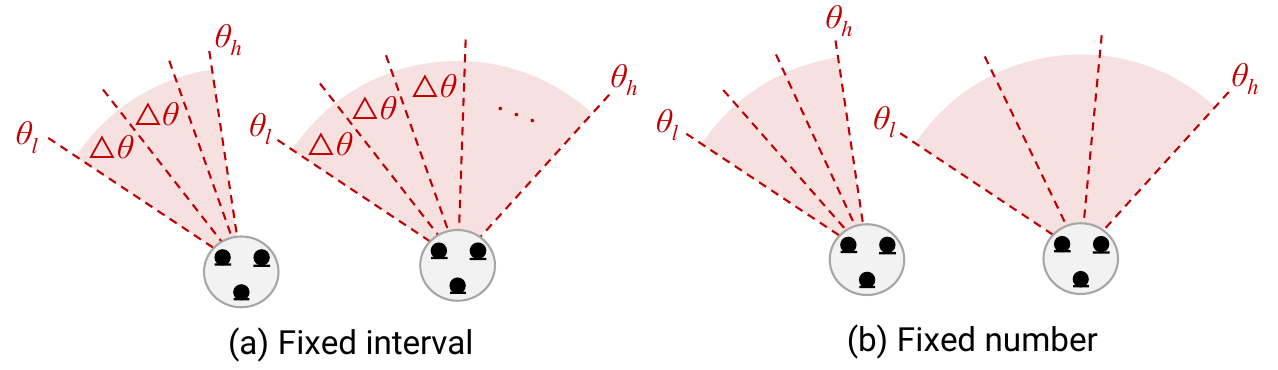}
\caption{The illustration of sampling in an azimuth window by (a) fixed interval or (b) fixed number.}
\label{fig:region_sample}
\end{figure}

The query region $\vec{r}$ contains angle windows while the definition of direction feature is based on discrete directions. A sampling process to sample certain directions within the angle window is thus necessary. Here we consider two types of sampling methods, which are shown in figure \ref{fig:region_sample}. For the sake of simplicity, we assume 1-D cases where only azimuth window $[\theta_l,\theta_h]$ is given, but it can easily be extended to 2-D cases where the elevation window $[\phi_l,\phi_h]$ is also provided: 
\begin{itemize}
    \item \textbf{Fixed interval}: The azimuth window is divided by a fixed pre-set interval $\Delta \theta$, which can be determined according to the spatial resolution of the microphone array. In this case, the number of spatial views $N=\lfloor \frac{\theta_h-\theta_l}{\Delta \theta}\rfloor+1$ will be varied from samples with different azimuth window widths. Figure~\ref{fig:region_sample} (a) shows this method.
    
    \item \textbf{Fixed number}: The azimuth window is evenly divided by a fixed number $N$ of spatial views irrelevant to the spatial resolution of the microphone array or the width of the window, where the $n$-th sampled azimuth is $\theta_n = \theta_l+(n-1)\frac{\theta_h-\theta_l}{N-1} (1\leq n\leq N)$. Figure~\ref{fig:region_sample} (b) shows this method.
\end{itemize}
Section~\ref{sec:ablation_feature} empirically compares the two schemes. For distance feature, there is no need to sample within a distance range, as the target sources within $[d_l,d_h]$ can be obtained by subtracting system outputs for query $[0,d_h]$ and $[0,d_l]$.

\subsection{Region feature aggregation}
\label{subsubsec:region_agg}

\begin{figure}[!t]
  \centering
  \includegraphics[width=\linewidth]{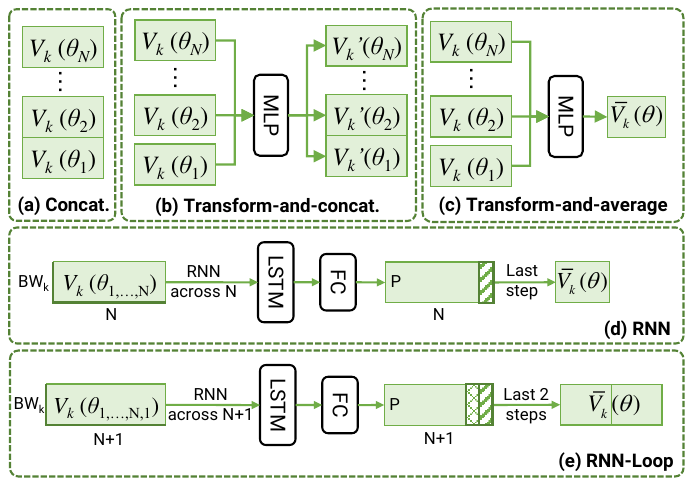}
\caption{Demonstration of different direction feature aggregation methods. The $n$-th direction feature at subband $k$ is denoted as $V_k(\theta_n)$ where $\textit{BW}_k$ corresponds to its bandwidth. The frame index $t$ is omitted.}
\label{fig:region_agg}
\end{figure}

\begin{figure*}[!t]
  \centering
  \includegraphics[width=\linewidth]{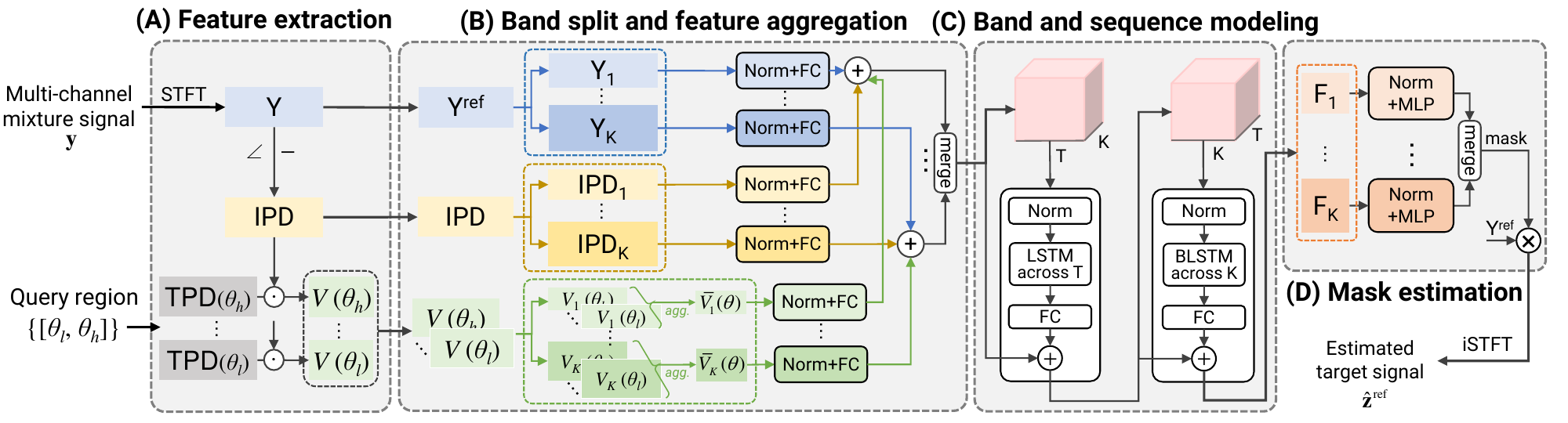}
\caption{The flowchart of A-ReZero where the query region is an angle window.}
\label{fig:a_rezero}
\end{figure*}

\begin{figure}[!ht]
  \centering
  \includegraphics[width=\linewidth]{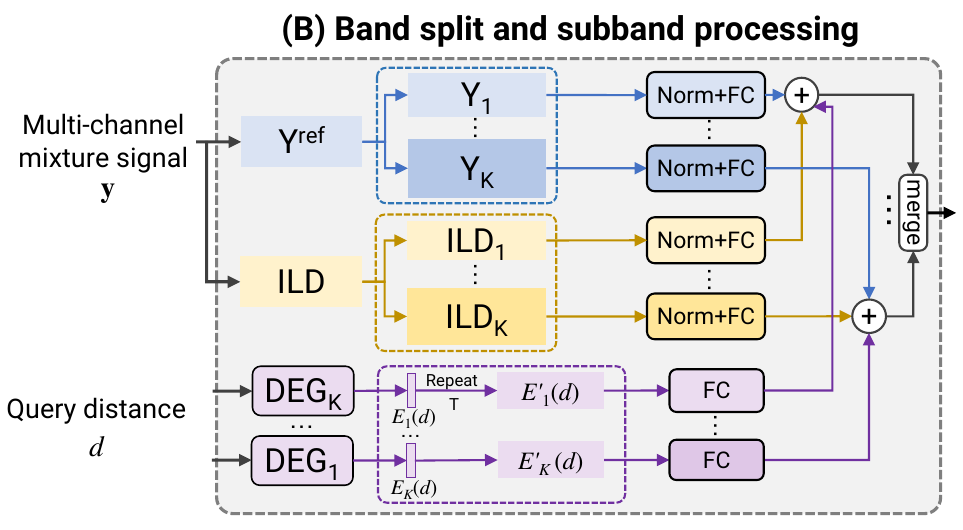}
\caption{The flowchart of band split and subband processing module in D-ReZero where the query region is a distance threshold.}
\label{fig:d_rezero}
\end{figure}

We aggregate the direction features sampled within the angle window to form a region descriptor for each frame. Here we formulate the feature aggregation operation at subband level, which is mainly inspired by recent advances in neural network designs for speech enhancement, such as NB-LSTM \cite{quan2022multi}, TF-gridnet \cite{wang2022tf}, band-split RNN \cite{luo2022bsrnn}, and Tea-PSE \cite{ju2022tea,ju2022tea2,ju2023tea}. Without loss of generality, we split the spectrogram and the corresponding direction features to $K\geq1$ subbands, and denote the $n$-th direction feature at $k$-th subband as $V_k(\theta_n) \in \mathbb{R}^{T\times \textit{BW}_k}$ where $\textit{BW}_k$ corresponds to its bandwidth. Note that by setting $K=1$ we obtain features for full bandwidth. We consider five types of feature aggregation methods: 
\begin{itemize}
    \item \textbf{Concatenate}: This corresponds to the most simple and straightforward method as the $N$ direction features $\{V_k(\theta_n, t)\}_{n=1}^N$ are directly concatenated along the bandwidth dimension to form the region descriptor $\overline{V}_k(\theta, t) \in \mathbb{R}^{N\cdot\textit{BW}_k}$. Figure~\ref{fig:region_agg} (a) shows this method.
    
    \item \textbf{Transform-and-Concatenate} (TAC): Inspired by \cite{luo2020end}, each sampled direction feature $V_k(\theta_n, t)$ is first transformed by an MLP shared by all features at the current subband. The MLP is applied to the bandwidth dimension of each feature to map it to another $P$-dimensional feature $\Bar{V}_k(\theta_n, t) \in \mathbb{R}^{N\cdot P}$. The region descriptor is then obtained by concatenating $\{\Bar{V}_k(\theta_n, t)\}_{n=1}^N$ along the feature dimension to form the region descriptor $\overline{V}_k(\theta, t) \in \mathbb{R}^{N\cdot P}$. Figure~\ref{fig:region_agg} (b) shows this method.

    \item \textbf{Transform-and-Average} (TAA): Similar to TAC, the sampled direction features are transformed by a shared MLP. The difference is that the region descriptor $\overline{V}_k(\theta, t) \in \mathbb{R}^{P}$ is obtained by averaging the transformed features $\{\Bar{V}_k(\theta_n, t)\}_{n=1}^N$. Figure~\ref{fig:region_agg} (c) shows this method.

    \item \textbf{RNN}: We first sort the $N$ sampled direction features with respect to their TDOAs and treat them as a feature sequence of length $N$. The sequence is then passed to a uni-directional long-short time memory (LSTM) layer \cite{hochreiter1997long} with hidden size $P$. The last step of the LSTM output is used as the region descriptor $\Bar{V}_k(\theta, t) \in \mathbb{R}^{P}$. Figure~\ref{fig:region_agg} (d) shows this method.
    
    \item \textbf{RNN-Loop}: Instead of using the sequence of length $N$, we further append the feature with the smallest TDOA, i.e., the first feature in the sequence, to the end of the sequence to form a ``closed-loop''. The feature sequence of length $N+1$ is then sent to the LSTM layer, and we use the concatenation of the last two steps of the LSTM outputs as the region descriptor $\Bar{V}_k(\theta, t) \in \mathbb{R}^{2P}$. Figure~\ref{fig:region_agg} (e) shows this method.
\end{itemize}
Section~\ref{sec:ablation_feature} empirically compares all the aforementioned methods. For distance feature, we apply subband-specific DEG modules to generate subband distance embeddings $\{E_k(d)\}_{k=1}^K$, and no feature aggregation operation is needed in this case.



\subsection{Multi-channel BSRNN}
\label{subsec:mch_bsrnn}

We also propose a neural network design that can better utilize the region descriptors to obtain a better source extraction performance. Inspired by the recent success of band-split RNN (BSRNN), here we extend the original BSRNN to the R-SE task. Figure~\ref{fig:a_rezero} shows the flowchart for the modified BSRNN architecture for angular region query, which we refer to as the A-ReZero model (angle-ReZero), for angle window query, which includes a feature extraction module, a band split and subband processing module, a band and sequence modeling module, and a mask estimation module.
\begin{itemize}
    \item \textbf{Feature extraction}: The complex spectrograms of $\{\vec{y}^m\}_{m=1}^M$ are first extracted by short-time Fourier transform (STFT) as the spectral feature, and the IPD, TPD and direction features are extracted accordingly.
    \item \textbf{Band split and feature aggregation}: A core design paradigm for BSRNN is its band-split operation. We split the complex spectrogram at a selected reference microphone $Y^{\text{ref}}(t, f)$, IPD and direction features into $K$ nonoverlapped subbands as described in section~\ref{subsubsec:region_agg}, and aggregate the direction features into a region descriptor by one of the feature aggregation methods. The real and imaginary parts of the complex subband spectrograms are concatenated to form a real-valued spectral feature. Each of the three features is then passed to a subband-specific batch normalization module \cite{ioffe2015batch} followed by a fully-connected (FC) layer to map to a same faeture dimension, and the outputs at each subband are summed to form the overall subband-level feature.
    \item \textbf{Band and sequence modeling}: This part is identical to the original BSRNN model, where each BSRNN block contains two residual RNN layers sequentially applied across the temporal and subband dimensions. To support streaming processing, we change the the bidirectional LSTM layer in the sequence modeling RNN to a uni-directional LSTM layer, and the layer normalization module in it to batch normalization module. The band modeling RNN is kept unchanged.
    \item \textbf{Mask estimation}: This is also the same to the original BSRNN model where subband-specific batch normalization modules and MLPs are used to estimate the complex-valued T-F masks for the reference microphone. The masked subband spectrograms are finally concatenated and reconstructed to waveform by inverse STFT operation.
\end{itemize}
We refer the interested readers to \cite{luo2023music} for more details on the BSRNN architecture. 

For distance threshold query, we modify the band split and feature aggregation module to accept the subband distance embeddings. Figure~\ref{fig:d_rezero} shows the modification from A-ReZero model to the D-ReZero model (distance-ReZero), where the ILD feature is calculated at subband level, and subband distance embeddings are repeated across the temporal dimension as it is time-invariant. The feature aggregation module is no longer required as no feature sampling is needed in this case. The subband spectral and ILD features are normalized and transformed in the same way as A-ReZero, while we remove the batch normalization operation for the distance embeddings.

\section{Experiment configurations}
\label{sec:config}
\subsection{Data preparation}

\begin{figure*}[!ht]
  \centering
  \includegraphics[width=16cm]{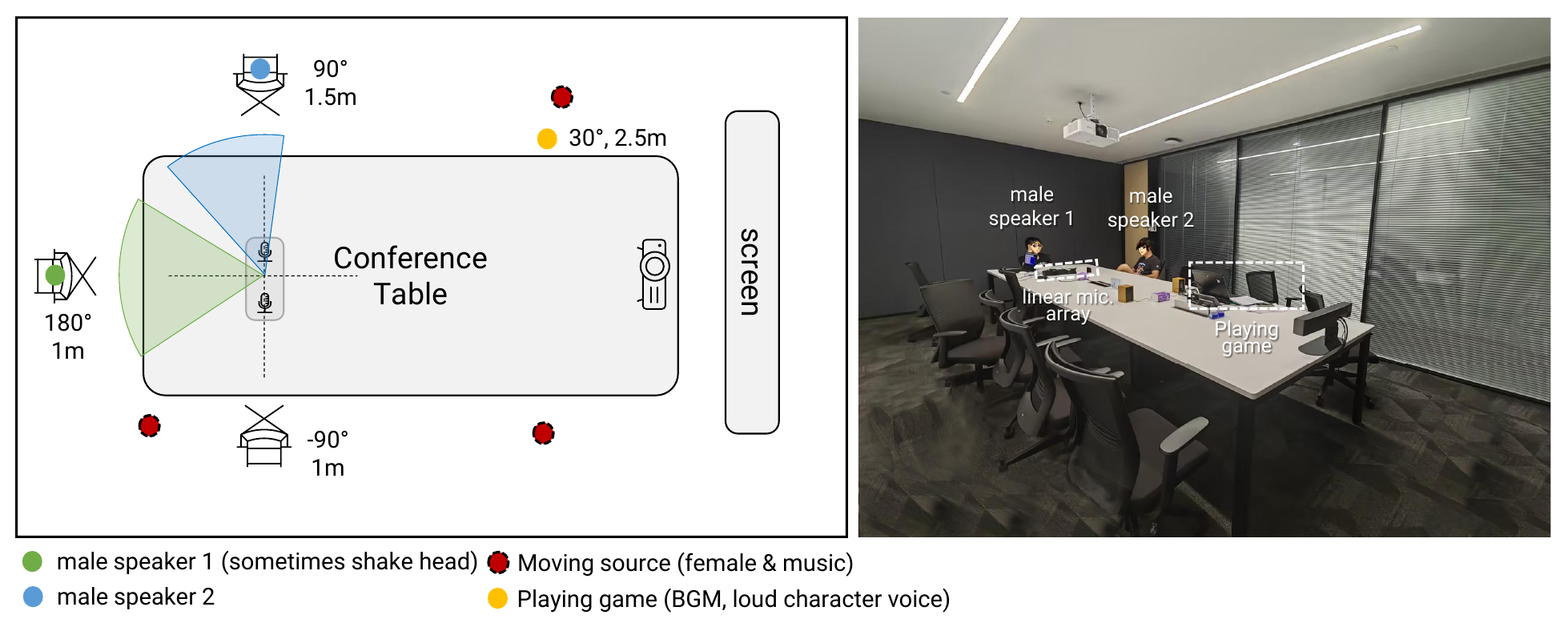}
  \caption{The conference room layout and the microphone array, speaker and interference locations for the real-recorded data.}
  \label{fig:conference_room}
\end{figure*}

Although ReZero can be applied to a wide range of microphone array geometries and task configurations, here we consider a typical meeting room scenario where a small-scale microphone array is put on a table and multiple speakers are seated around the table, and the task is defined to perform joint target region speech extraction and dereverberation. We assume two types of microphone array geometries, a circular array and a linear array, each with 8 microphones. The circular array has a diameter of 5~cm, and the microphones in the linear array are evenly distributed within a 22.5 cm diameter. In this configuration, we consider query region types of 1-D angular (azimuth windows), spherical and conical, as elevation cue is typically not as important as azimuth cue in such meeting scenarios when the microphone array is placed on the table. We put additional data simulation configurations and results for 2-D angular query region in additional materials\footnote{\url{https://innerselfm.github.io/rezero/}}. 

The training data generation, including room impulse response (RIR) simulation, random query region selection, and dynamic mixing \cite{subakan2021attention}, is performed completely on-the-fly. A mixture signal is generated by randomly sampling [1, 2] speech signals from the \emph{train-clean-100} subset of LibriSpeech corpus \cite{panayotov2015librispeech} and [1, 4] noise signals from the combination of \emph{100 Nonspeech corpus} \cite{hu2010nonspeech}, \emph{MUSAN corpus} \cite{snyder2015musan}, 1000 internal instrumental music pieces, and isotropic noise generator \cite{habets2007isotropic}, all with 4 to 6 second length. The multi-channel room impulse responses (RIRs) are simulated using the fast random simulation of multi-channel RIR (FRAM-RIR) method \cite{luogu2023framrir}. The positions of microphones and speech and noise signals are randomly sampled within a randomly generated room whose dimensions varying from $3\times3\times2.5$ meters to $10\times8\times4$ meters. The signal locations are constrained to have a minimum distance of 0.5 meters from the walls. The reverberation time (T60) ranges from 0.05 to 0.7 seconds. For dereverberation, we set the direct and early reflection component as the training target, where the early reflection context is set as [-6, 50] ms around the first peak of the direct path RIR. The speech and noise signals, excluding the isotropic noise signals, are convolved with their corresponding RIR filters. After that, the signal-to-interference ratios (SIRs) between the first sampled speech signal and other speech signals, if any, are randomly sampled within -6 and 6 dB. The SIRs between the first sampled noise signal and other noise signals, if any, are randomly sampled within -15 and 15 dB. The signal-to-noise ratio (SNR) between the sum of all speech signals and the sum of all noise signals is randomly sampled within 5 and 15 dB. All signals are sampled at 16 kHz. The azimuth range of the query region $[\theta_l,\theta_h]$ is randomly set with the constraint $30\degree \le \theta_h-\theta_l \le 90\degree$. The distance threshold of the query region is set within $[0.2, 2.0]$ meters. The proportions of utterance with $Q=0$, $Q=1$ and $Q=2$ during on-the-fly training are about 27\%, 65\% and 8\% for angular query region, and 10\%, 45\% and 45\% for both spherical and conical query regions, respectively. 

For simulated evaluation data, we generate 3000 mixture utterances ($\sim$4.2 hours) using \emph{train-clean-360} split of LibriSpeech. To validate the generalization capability of the proposed model for possibly mismatched room acoustics, we use gpuRIR \cite{diaz2020gpurir} as the RIR simulator. The azimuth and elevation ranges and the distance thresholds for the query region of each mixture are sampled within the same constraints as the training configurations per mixture. The proportions of utterance with $Q=0$, $Q=1$ and $Q=2$ in the evaluation data is 28\%, 36\% and 36\% for angular query region, 14\%, 33\% and 53\% for spherical query region, and 33\%, 36\% and 31\% for conical query region, respectively. For real-recorded evaluation data, we record 5-minute-long natural conversation sessions using a 8-mic linear array (with 3.5 cm spacing each) in a conference room. The layout of the conference room is illustrated in Figure \ref{fig:conference_room}. The room size is about $5 \times 8\times 3$ meters. 
A typical session is that there are two male speakers (the green and blue points) sitting on the chair and having a casual conversation, while another person (the yellow point) is sitting at 30$\degree$ relative to the microphone array and playing mobile game. There is also another person wandering in the room while holding a mobile phone playing music (the red dashed points). Results for the real-recorded data is available online\footnote{\url{https://innerselfm.github.io/rezero/}}.

\begin{table*}[!ht]
  \caption{R-SE results of different region sampling strategies. 
  The result of each configuration is obtained by repeating the experiment 3 times and expressed as the mean$\pm$standard deviation (\textit{mean$\pm$std}).}
  \label{tab:ablation1_region_sampling_full}
  \centering
  \small
  \scalebox{1.0}{
  \begin{tabular}{l|c|cc|c|c}
    \toprule
    \multirow{2}{*}{\textbf{Region sampling}} &
    \textbf{Decay} (dB) &
    \multicolumn{2}{c|}{\textbf{SDR} (dB)} &    
    \multirow{2}{*}{\textbf{STOI} (\%)} &
    \multirow{2}{*}{\textbf{PESQ}} \\
    & Q=0 & Q=1 & Q=2 & \\
    \midrule
    Mixture & -- & 0.0 & 9.63 & 70.4 & 1.18  \\
    \hline
    Fixed interval = 10$\degree$  & 49.44$\pm$2.77 & 12.36$\pm$0.10 & 13.47$\pm$0.47  & 90.0$\pm$0.7 & 2.22$\pm$0.03\\
    Fixed interval = 15$\degree$  & 47.45$\pm$1.36 & \textbf{12.78}$\pm$0.50 & \textbf{14.13}$\pm$0.20 & \textbf{91.2}$\pm$0.2 & 2.27$\pm$0.02\\
    Fixed interval = 20$\degree$  & 45.06$\pm$1.95 & 12.38$\pm$0.12 & 13.46$\pm$0.83 & 90.3$\pm$0.8 & 2.21$\pm$0.04 \\
    \hline
    Fixed number = 3  & 48.69$\pm$2.82 & 12.39$\pm$0.21 & 13.84$\pm$0.31 & 90.9$\pm$0.4 & \textbf{2.28}$\pm$0.05  \\
    Fixed number = 4  & \textbf{52.16}$\pm$2.58 & 12.36$\pm$0.27 & 13.34$\pm$0.33 & 90.4$\pm$0.5 & 2.24$\pm$0.03 \\
    Fixed number = 6  & 49.42$\pm$4.17 & 12.35$\pm$0.10 & 13.87$\pm$0.30 & 90.9$\pm$0.3 & 2.25$\pm$0.01 \\
    Fixed number = 8  & 52.03$\pm$4.89 & 12.39$\pm$0.09 & 13.74$\pm$0.18 & 90.9$\pm$0.2 & \textbf{2.28}$\pm$0.01 \\
    \bottomrule
  \end{tabular}
  }
\end{table*}

\begin{table*}[!ht]
  \caption{R-SE results of different region feature aggregation methods. Number / 8 is used as the region sampling method.}
  \label{tab:ablation2_region_aggregation_full}
  \centering
  \small
  \begin{tabular}{l|c|cc|c|c}
    \toprule
    \multirow{2}{*}{\textbf{Region aggregation}} &
    \textbf{Decay} (dB) &
    \multicolumn{2}{c|}{\textbf{SDR} (dB)} &    
    \multirow{2}{*}{\textbf{STOI} (\%)} &
    \multirow{2}{*}{\textbf{PESQ}} \\
    & Q=0 & Q=1 & Q=2 & \\
    \midrule
    Mixture & -- & 0.0 & 9.63 & 70.4 & 1.18 \\
    \hline
    Concatenate  & 50.22$\pm$1.25 & 12.41$\pm$0.08 & 13.55$\pm$0.21  & 90.8$\pm$0.2 & 2.57$\pm$0.03\\
    Transform-and-Concatenate (TAC)   & 44.62$\pm$5.35 & 11.24$\pm$0.27 & 12.58$\pm$0.42  & 90.6$\pm$0.4 & 2.48$\pm$0.02\\
    Transform-and-Average (TAA)  & 39.27$\pm$1.47 & 11.28$\pm$0.22  & 10.99$\pm$0.29 & 90.0$\pm$0.6 & 2.30$\pm$0.13  \\ 
    RNN  & 52.03$\pm$4.89 & 12.39$\pm$0.09 & 13.74$\pm$0.18  & \textbf{90.9}$\pm$0.2 & 2.58$\pm$0.01  \\
    RNN-Loop  & \textbf{53.03}$\pm$2.37 & \textbf{12.48}$\pm$0.24 & \textbf{13.88}$\pm$0.18 & 90.9$\pm$0.4 & \textbf{2.59}$\pm$0.04 \\
    \bottomrule
  \end{tabular}
\end{table*}

\subsection{Model and training configurations}
\label{sec:train_config}
We use 32~ms window size and 8~ms hop size with Hann window for STFT for all experiments. We set the band-split scheme in the modified BSRNN model to be slightly different than the single-channel counterpart, where we split the spectrogram into ten 100 Hz bandwidth subbands, twelve 200 Hz bandwidth subbands, eight 500 Hz bandwidth subbands, and treat the rest as another subband. This results in 31 subbands. We set the number of band and sequence modeling modules to 8 and the feature dimension to 48. We set the feature dimension for the aggregated region features $P$ to 16, and we use all possible microphone pairs ($C^2_8=28$ with 8 microphones) to calculate the spatial and region features. We use the AdamW optimizer \cite{loshchilov2019adamw} with initial learning rate of $1e^{-3}$, and the learning rate is decayed by 0.98 for every two epochs. All models are trained for 240k iterations with a batch size of 8.



The training target varies as $Q$ varies in equation~\ref{eq:target}. When $Q=0$, the model is expected to generate a silent signal since there is no speech source within the target zone. When $0<Q<C$, the model performs joint denoising, source extraction and dereverberation. When $Q=C$, the model preserves all speakers while only performs denoising and dereverberation on them. The loss function is a combination of frequency domain mean absolute error (freq-MAE) and standard signal-to-noise ratio (SNR) \cite{wisdom2021s} and is dependent on $Q$:
\begin{equation}
    \label{eq:loss} 
    \mathcal{L} =
    \left\{ 
    \begin{array}{lr}
         \lambda \left( ||\mathcal{R}(\hat{\vec{Z}}^{\text{ref}}) ||_1 +  ||\mathcal{I}(\hat{\vec{Z}}^{\text{ref}}) ||_1 \right), & Q=0 \\
        \text{SNR}(\vec{z}^{\text{ref}}, \hat{\vec{z}}^{\text{ref}}), & Q>0
    \end{array} \right.
\end{equation}
where $\mathcal{R}({\hat{\vec{Z}}^{\text{ref}})}$ and $\mathcal{I}({\hat{\vec{Z}}^{\text{ref}})}$ are the real and imaginary parts of the spectrogram of the estimated target $\hat{\vec{z}}^{\text{ref}}$, respectively, and $\lambda$ is a weighting factor that we empirically set to 0.01.

\subsection{Evaluation metrics}
Different metrics are used for utterances where $Q>0$ and $Q=0$. We use signal-to-distortion ratio (SDR) \cite{le2019sdr} for utterances where $Q>0$, and additionaly we use short-term objective intelligibility (STOI) \cite{taal2010stoi} and wideband perceptual evaluation of speech quality (PESQ) \cite{rix2001pesq} for utterances where $Q=1$. when $Q=0$, we calculate the energy decay of the model output with respect to the mixture in decibel scale to measure how well the model estimates silence.


\section{Results and analysis}
\label{sec:result}
We start with the experiment results and ablation studies for angular query region with circular array, and then move to the results for linear array and spherical and conical query regions.

\subsection{Results for angular query region with circular array}
\subsubsection{Effect of region sampling and aggregation configurations}
\label{sec:ablation_feature}

We first examine the effect of different region sampling strategies described in section~\ref{subsubsec:region_sample}. Table~\ref{tab:ablation1_region_sampling_full} shows the performance of models trained with different region sampling strategies and configurations, where we use the ``RNN'' region feature aggregation method described in section~\ref{subsubsec:region_agg} for comparison. We can see that fixed-number-based region sampling strategy is in general on par with fixed-interval-based strategy with no significant differences, indicates that once the proper region features are calculated, the model is relatively insensible to how they are sampled. We thus select a fixed-number-based strategy with number of 8 as the default setting for other experiments, as it keeps the number of region features identical even when the angle window widths changes, and a larger number of samples can guarantee a fine-grained spatial resolution of the features.

We then compare different region feature aggregation methods in table~\ref{tab:ablation2_region_aggregation_full}. We observe that concatenation is a simple yet effective method and is consistently better than TAC and TAA in all scenarios. RNN and RNN-Loop are on par with concatenation, with the additional advantage that these two methods can handle any number of region features (e.g., number of spatial views). As RNN-Loop is slightly better than RNN, we thus select RNN-Loop as the default region feature aggregation method for the following experiments.

\begin{table*}[!ht]
  \caption{Comparison with oracle beamforming, oracle target speech separation methods and other SOTA speech enhancement methods. N/A: Not Applicable for this case.} 
  \label{tab:compare_full}
  \centering
  \small
  \scalebox{0.95}{
  \begin{tabular}{l|l|c|c|cc|c|c|c|c}
    \toprule
    \multirow{2}{*}{\textbf{Method}} &
    \multirow{2}{*}{\textbf{Oracle / Auxiliary information}} &
    \multirow{2}{*}{\textbf{Causal}} &    
    \textbf{Decay} (dB) &
    \multicolumn{2}{c|}{\textbf{SDR} (dB)} &    
    \multirow{2}{*}{\textbf{STOI} (\%)} &
    \multirow{2}{*}{\textbf{PESQ}} &
    \textbf{\#param.} &
    \textbf{MACs}  \\
    & & & Q=0 & Q=1 & Q=2 & & &(M) &(G/s)\\
    \midrule
    Mixture & -- & --  & -- & 0.00 & 9.63 & 70.4 & 1.18 & - & -\\
    \hline
    \multicolumn{5}{l}{\textbf{Oracle separation methods}} \\
    \hline
    B-SS-BSRNN & target source assignment, LBT & $\checkmark$ & N/A & 10.92 & 12.32 & 89.5 & 2.04 & 2.96 & 5.96  \\
    D-SE-BSRNN & target azimuth $\theta_q$ & $\checkmark$ & N/A & 11.96 & 7.00 & 90.5 & 2.19 & 2.87 & 5.94 \\
    D-SE-BSRNN & target azimuth $\theta_q$ with $\pm15\degree$ error & $\checkmark$ & N/A & 11.85 & 6.76 & 89.9 & 2.15 & 2.87 & 5.94 \\    
    \hline
    \multicolumn{5}{l}{\textbf{Oracle beamforming methods}} \\ 
    \hline
    IRM-MVDR & target IRM & × & \textbf{$>$60} & 8.31 & 6.93 & 90.3 & 2.08 & -- & -- \\
    CRM-MVDR & target complex spectrogram & × & \textbf{$>$60} & 5.95 & 12.43 & 81.5 & 1.41 & -- & -- \\
    DAS & target azimuth $\theta_q$ &$\checkmark$ & N/A & 0.14 & 8.02 & 73.1 & 1.21 & -- & -- \\    
    \hline
    \multicolumn{5}{l}{\textbf{Speech enhancement and extraction methods}} \\ 
    \hline 
    SE-BSRNN & -- &$\checkmark$ & N/A & 4.67 & 5.19& 71.6 & 1.59 & 13.6 & 18.2 \\  
    P-SE-BSRNN & target speaker enrolment ($>$4s) &$\checkmark$ & N/A & 5.65 & 5.57 & 83.6 & 1.72 & 22.2 & 14.7 \\    
    P-SE-BSRNN & target speaker enrolment ($>$4s) & × & N/A & 8.18 & 7.49 & 82.6 & 1.87 & 23.6 & 23.4 \\    
    \hline\hline
    A-ReZero & target angular region &$\checkmark$ & 53.03 & \textbf{12.48} & \textbf{13.88} & \textbf{90.9} & \textbf{2.29} & 3.00 & 6.03\\
    \bottomrule
  \end{tabular}
  }
\end{table*}

\subsubsection{Comparison with other methods}
As mentioned in section~\ref{sec:related}, there is no existing R-SE models with fully customizable region queries to the best of our knowledge, hence we compare ReZero with direction-based separation systems with oracle target source direction information, beamforming methods with oracle target source T-F masks, spectrograms or directions, and state-of-the-art (SOTA) speech enhancement and extraction methods with or without target speaker enrollment. To be specific, the benchmark methods we select are:
\begin{itemize}
    \item \textbf{Oracle separation methods}: We use BSRNN with the same model architecture we described in section~\ref{sec:train_config} with different types of oracle target source direction information. The blind source separation BSRNN (B-SS-BSRNN) model does not use any location information and performs blind separation of all available sources. All separated sources, as well as all of their possible combinations (for the cases where $Q>1$), are treated as possible model outputs, and the one with the highest SDR with the target source in the query region is selected. This is similar to the evaluation of existing B-SS systems with oracle source assignment. We adopt location-based training (LBT) \cite{taherian2022multi}, which sorts the training target by their locations (azimuth in this case), instead of permutation-invariant training (PIT) \cite{yu2017permutation} during training phase. The direction-informed speech extraction BSRNN (D-SE-BSRNN) replaces the sampled region features by a single target direction feature using oracle target direction and removed the region feature aggregation module, and all other components are kept identical to ReZero. For the case of $Q=2$, we run D-SE-BSRNN twice with two target source directions and sum the outputs. We also add a random $\pm15\degree$ perturbation to the target source direction to evaluate the effect of inaccurate direction information.
    \item \textbf{Oracle beamforming methods}: We select oracle beamforming methods including ideal ratio mask based minimum variance distortionless response (MVDR) beamformer (IRM-MVDR) \cite{heymann2016neural}, complex spectral mapping based MVDR (CSM-MVDR) \cite{tan2022neural,wang2020complex}, and delay-and-sum (DAS) beamformer. For IRM-MVDR and CSM-MVDR, we use the oracle IRM or complex spectrogram for the target source (sum of all active speakers when $Q=2$), and for DAS, we run the beamforming process twice similar to D-SE-BSRNN.
    \item \textbf{Speech enhancement and extraction methods}: We select strong benchmarks for single-channel speech enhancement and extraction methods which were ranked top 3 in the 5th deep noise suppression (DNS) challenge \cite{yu2023dns}. The models are all BSRNN-based with different model configurations, and details about the models can be found in \cite{yu2022high, yu2023dns}. The speech enhancement BSRNN (SE-BSRNN) does not have the ability to distinguish region queries and can be viewed as a single-channel speech enhancement baseline. The personalized speech extraction BSRNN (P-SE-BSRNN) takes an additional target speaker enrollment to perform target speaker extraction, and again we run the model twice if there are more than one speakers in the query region.
\end{itemize}

\begin{table*}[!ht]
  \caption{Performance comparison of models with different numbers of microphones or different microphone array diameters in circular microphone array.}
  \label{tab:ablation4_nonlinear_ma_full}
  \small
  \centering
  \scalebox{1.0}{
  \begin{tabular}{l|c|cc|c|c}
    \toprule
    \small
    \multirow{2}{*}{\textbf{mic. config}} &
    \textbf{Decay} (dB) &
    \multicolumn{2}{c|}{\textbf{SDR} (dB)} &    
    \multirow{2}{*}{\textbf{STOI} (\%)} &
    \multirow{2}{*}{\textbf{PESQ}} \\
    & Q=0 & Q=1 & Q=2 & & \\
    \midrule
    \multicolumn{6}{l}{\text{Diameter = 5 cm}} \\
    \hline
    3 mic & 48.38$\pm$2.42 & 11.92$\pm$0.12 & 13.16$\pm$0.04 &89.8$\pm$0.1 & 2.12$\pm$0.01\\

    4 mic  & 49.55$\pm$1.12 & 12.35$\pm$0.07 & 13.79$\pm$0.18 & 90.8$\pm$0.4 & 2.25$\pm$0.01 \\
    
    6 mic & 49.32$\pm$0.41 & 12.37$\pm$0.04 & 13.80$\pm$0.12 & 90.8$\pm$0.1 & 2.24$\pm$0.02  \\

    8 mic  & 52.03$\pm$4.89 & 12.39$\pm$0.09 & 13.74$\pm$0.18 & 90.9$\pm$0.2 & 2.28$\pm$0.01 \\

    \hline
    \multicolumn{6}{l}{\text{\#mic = 8}} \\
    \hline
    $d=15$ cm & 53.06$\pm$1.13  & 13.29$\pm$0.12 & 14.60$\pm$0.25 & 91.7$\pm$0.2 & 2.36$\pm$0.03 \\
    $d=10$ cm  & 53.13$\pm$0.37 & 12.94$\pm$0.08 & 14.48$\pm$0.08 & 91.3$\pm$0.1 & 2.39$\pm$0.00\\
    $d=7$ cm & 52.19$\pm$2.09 & 12.65$\pm$0.02 & 14.48$\pm$0.09 & 91.6$\pm$0.1 & 2.37$\pm$0.02\\
    $d=5$ cm  & 52.03$\pm$4.89 & 12.39$\pm$0.09 & 13.74$\pm$0.18 & 90.9$\pm$0.2 & 2.28$\pm$0.01  \\
    \bottomrule
  \end{tabular}
  }
\end{table*}

\begin{table}[!t]
  \caption{R-SE results of different numbers of elements and diameters of the linear microphone array. Number / 8 and RNN-L is used as the region sampling method and region aggregation method, respectively.}
  \label{tab:ablation4_linear_ma_full}
  \small
  \centering
  \scalebox{0.99}{
  \begin{tabular}{l|c|cc|c|c}
    \toprule
    \small
    \multirow{2}{*}{\textbf{\#mic.}} &
    \textbf{Decay} (dB) &
    \multicolumn{2}{c|}{\textbf{SDR} (dB)} &    
    \multirow{2}{*}{\textbf{STOI} (\%)} &
    \multirow{2}{*}{\textbf{PESQ}} \\
    & Q=0 & Q=1 & Q=2 & \\
    \midrule
    \multicolumn{6}{l}{\text{Diameter = 22.5 cm}} \\
    \hline
    Mixture & -- & 0.19 & 9.41 & 70.8 & 1.21 \\
    \hline
    2 mic & 37.16 & 10.82 & 12.66 & 86.9 & 1.97 \\

    4 mic  & 39.77 & 11.52 & 13.64 & 89.4 & 2.13\\
    
    8 mic  & 42.90 & 11.44 & 13.35 & 89.7 & 2.10 \\
    \bottomrule
  \end{tabular}
  }
\end{table}

Table \ref{tab:compare_full} shows the comparison between the proposed A-ReZero model and other benchmark systems. We can observe that oracle separation methods and speech enhancement and extraction methods cannot handle cases where $Q=0$, as they cannot take any region feature into account. Oracle IRM-MVDR and CSM-MVDR can completely cancel all signals in this case, as the oracle IRM or spectrogram is all-zero in this case. DAS does not have the target location information and cannot generate silent output either. A-ReZero is able to achieve comparable energy suppression ability as oracle IRM-MVDR and CSM-MVDR in this case. For $Q>0$, oracle separation methods are in general better than other benchmark methods, while the proposed A-ReZero method still outperforms all methods in terms of SDR and PESQ and is on par with IRM-MVDR in terms of STOI. Moreover, the model size and complexity of A-ReZero are on par with oracle separation methods and much smaller and lower than speech enhancement and extraction methods, which indicates that the use of proper region feature calculation and aggregation methods is beneficial to the model performance without drastically increasing the model complexity.

\subsubsection{Ablation study on microphone geometry configurations} 

We then investigate the effect of different microphone geometry configurations. Table~\ref{tab:ablation4_nonlinear_ma_full} shows the model performance with different numbers of microphones and array diameters in the circular microphone array. For different numbers of microphones, we select a subset of the 8 microphones and adjust the number of microphone pairs accordingly (remember that with RNN-Loop feature aggregation method, the model is insensible to the number of microphone pairs). For different diameters we train the models with corresponding configurations and generate test data with identical source selections and locations, query regions, room acoustics but only change the array diameter for a fair comparison. We observe that the performance of 3 microphones is worse than that of 4 microphones in all conditions, and the model performance with 4 microphones is comparable to that with 6 or 8 microphones when $Q>0$ and slightly worse when $Q=0$. It implies that it might not be necessary to use the spatial and region features from all microphone pairs when there is adequate number of microphones. We also find that increasing the microphone array diameter can lead to consistent performance improvement, which indicates that the spatial resolution of the microphone array, which affects the spatial and region features, can be important in the R-SE task.

We now investigate whether the aforementioned observations are sill valid for linear microphone array. Table~\ref{tab:ablation4_linear_ma_full} shows the performance of the linear array with different numbers of microphones. We observe that using more than 2 microphones is still beneficial, while the performance with 4 microphones is also on par or even slightly better than that with 8 microphones. This matches the observations in circular array. Moreover, comparing the absolute values for SDR, STOI and PESQ in table~\ref{tab:ablation4_nonlinear_ma_full} and table~\ref{tab:ablation4_linear_ma_full}, we find that the overall performance in linear array is worse than that in circular array. One possible reason is that linear arrays are not able to distinguish sources located symmetrically around the array (i.e., with same TDOA to all microphones), hence it is possible that symmetric target sources are leaked into the query region and hurt the performance.

\subsection{Results on spherical region query}

\begin{table}[!t]
  \caption{Results with spherical query regions.} 
  \label{tab:sphere}
  \centering
  \small
  \scalebox{0.9}{
  \begin{tabular}{l|c|cc|c|c}
    \toprule
    \multirow{2}{*}{\textbf{Method}} & 
    \textbf{Decay} (dB) &
    \multicolumn{2}{c|}{\textbf{SDR} (dB)} &    
    \multirow{2}{*}{\textbf{STOI} (\%)} &
    \multirow{2}{*}{\textbf{PESQ}}  \\
    & Q=0 & Q=1 & Q=2 & & \\
    \midrule
    Mixture & -- & 0.57 & 9.52 & 74.0 & 1.24  \\
    \hline
    D-ReZero & 32.88 & 10.59 & 14.88 & 89.9 & 2.38 \\
    \bottomrule
  \end{tabular}}
\end{table}

\begin{figure}
    \centering
    \includegraphics[width=\linewidth]{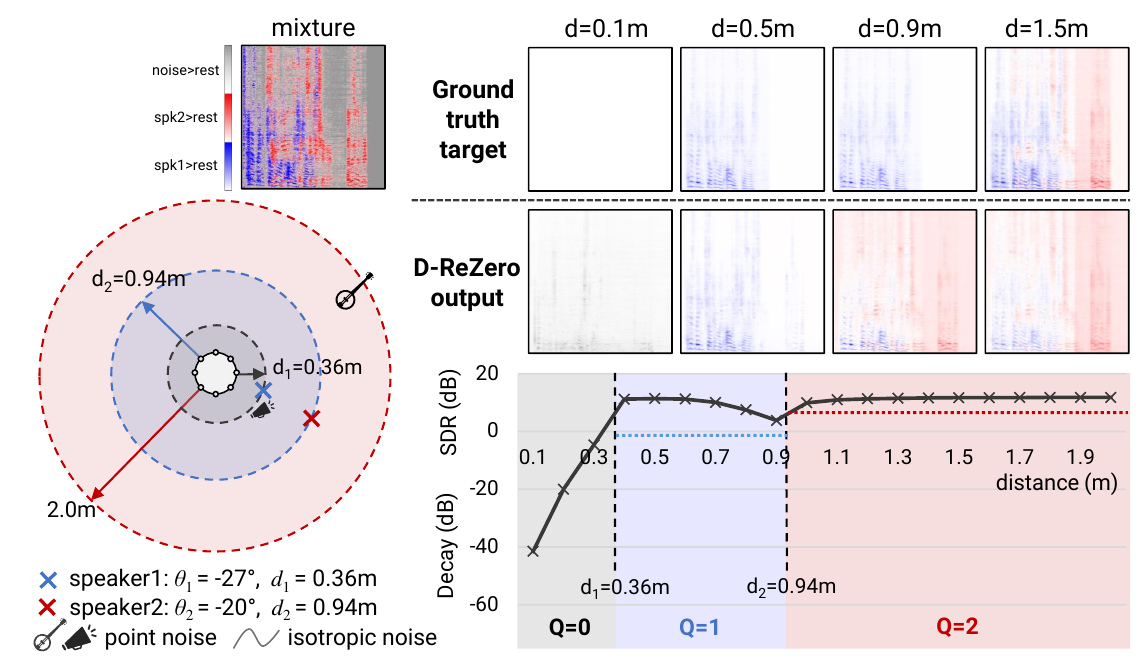}
    \caption{Black line: The energy decay ($Q=0$) or SDR ($Q>0$) results with respect to different query distances. Blue dotted line: The SDR of mixture signal using $Q=1$ target as reference signal. Red dotted line: The SDR of mixture signal using $Q=2$ target as reference signal. }
    \label{fig:s_rezero_example}
\end{figure}

We now move to spherical query region to test the model's performance with distance thresholds. Here we use the standard 8-microphone array with 5 cm diameter in the experiments above. Table~\ref{tab:sphere} shows the performance of D-ReZero and figure~\ref{fig:s_rezero_example} provides an example of the model's behavior with different distance thresholds. We assign random distance thresholds within $[0.2, 2.0]$ meters with a resolution of 0.1 meters during evaluation. Since there is no existing baseline models for adjustable distance threshold queries, here we simply present the performance of our D-ReZero model. We can see from figure~\ref{fig:s_rezero_example} that the model learns to smoothly transfer from silent outputs ($Q=0$) to extract the first speaker ($Q=1$) before the distance threshold reaches 0.4 meters, and it starts extracting the second speaker ($Q=2$) before the distance threshold reaches the actual distance of the second speaker (0.9 meters). This example, together with the performance shown in table~\ref{tab:sphere}, proves the effectiveness of our proposed DEG module as well as the whole D-ReZero model.

\begin{table}[!ht]
  \caption{R-SE results for the task of cone region sound extraction. } 
  \label{tab:cone}
  \centering
  \scalebox{0.85}{
  \begin{tabular}{l|c|cc|c|c}
    \toprule
    \multirow{2}{*}{\textbf{Method}} &    
    \textbf{Decay} (dB)&
    \multicolumn{2}{c|}{\textbf{SDR} (dB)} &    
    \multirow{2}{*}{\textbf{STOI} (\%)} &
    \multirow{2}{*}{\textbf{PESQ}}  \\
     & Q=0 & Q=1 & Q=2 & & \\
    \midrule
    Mixture & -- & -0.15 & 9.75 & 70.1 & 1.17  \\
    \hline
    A$\cap$D-ReZero & 50.16 & 11.29 & 11.46 & 85.9 & 2.07 \\    
    D$\rightarrow$A-ReZero & \textbf{75.96} & 12.25 & 14.06 & 88.2 & 2.13 \\
    A$\rightarrow$D-ReZero & 67.96 & \textbf{12.47} & \textbf{15.58} & \textbf{91.2} & \textbf{2.34} \\
    \bottomrule
  \end{tabular}}
\end{table}

\subsection{Results on conical region query}

\begin{figure}[!t]
    \centering
    \includegraphics[width=7.5cm]{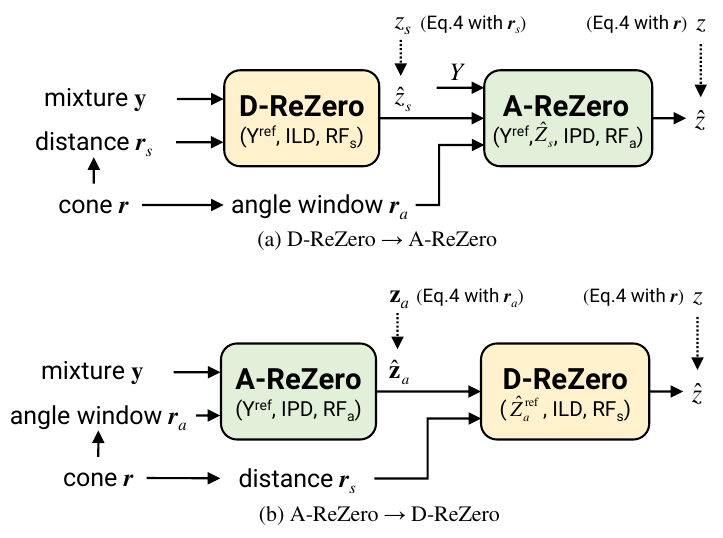}
    \caption{The flowcharts of (a) D-ReZero $\rightarrow$ A-ReZero and (b) A-ReZero $\rightarrow$ D-ReZero for conical region query. $\text{RF}_s$ and $\text{RF}_a$ indicate the distance query descriptor and direction query descriptor, respectively.}
    \label{fig:c_rezero}
\end{figure}

We then provide results and an example for conical region query. Although it is possible to train a model to take both direction and distance features as input features, we empirically found that the direction feature is too strong and the DEG module failed to be properly trained to let the distance feature take effect. Hence we investigate three alternative ways by using A-ReZero and D-ReZero models:


\begin{itemize}
    \item \textbf{A$\cap$D-ReZero}: The most simple way is to use pretrained A-ReZero and D-ReZero models to separately take the direction and distance features in the conical query region to obtain two outputs, and then calculate their T-F bin level intersection by selecting the T-F bin with smaller energy. 
    \item \textbf{D-ReZero $\rightarrow$ A-ReZero}: We first use a D-ReZero model to extract the output within the given distance threshold, and then we use an A-ReZero model on the output to further extract the output within the given angle window. Figure~\ref{fig:c_rezero} (a) shows the pipeline of this model combination scheme. 
    \item \textbf{A-ReZero $\rightarrow$ D-ReZero}: We first use an A-ReZero model to extract the output within the given angle window, and then we use a D-ReZero model to extract the output within the given distance threshold. Figure~\ref{fig:c_rezero} (b) shows the pipeline of this model combination scheme. 
\end{itemize}
Both D-ReZero $\rightarrow$ A-ReZero and A-ReZero $\rightarrow$ D-ReZero are trained from scratch with training objectives applied to A-ReZero and D-ReZero sub-models separately. Table~\ref{tab:cone} shows the performance of the three model combination methods, and we can see that the A-ReZero $\rightarrow$ D-ReZero scheme performs better than the other two schemes. We thus set A-ReZero $\rightarrow$ D-ReZero as the default configuration for conical region queries.

\begin{figure*}[!t]
    \centering
    \includegraphics[width=\linewidth]{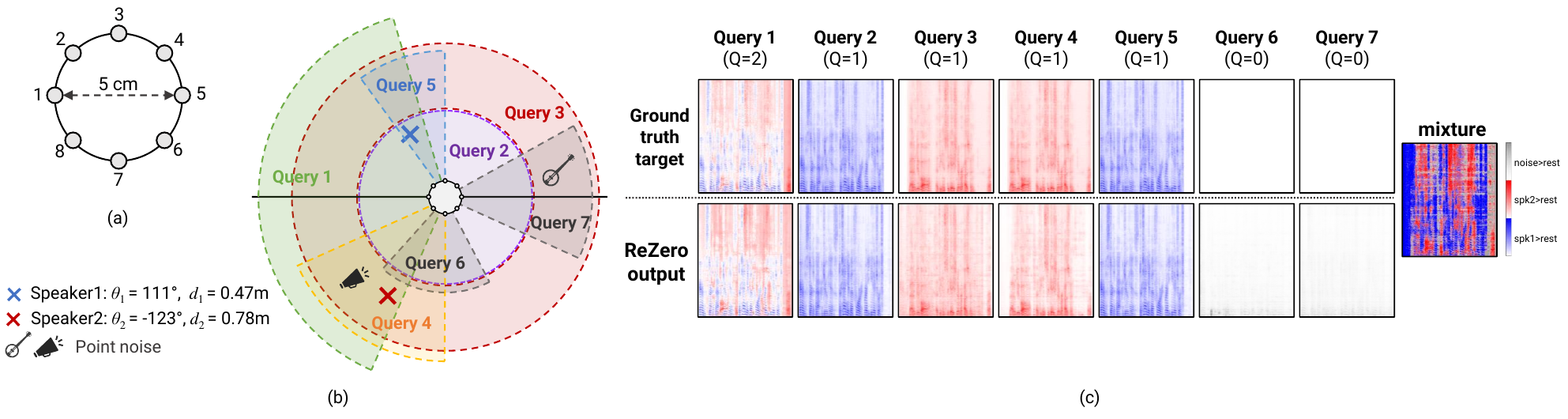}
    \caption{(a) The 8-microphone circular microphone array with diameter of 5 cm. (b) The illustration of different query regions on the same utterance. (c) The ground truth targets and ReZero outputs of different query regions. }
    \label{fig:rezero_example}
\end{figure*}


\subsection{Example on alterable region queries}

As a final remark, we provide an example on alternating different types of region queries on a same utterance. Figure~\ref{fig:rezero_example} shows the source locations and region query configurations, where there are two speakers and two point noises and seven different types of region queries are evaluated. Table~\ref{tab:cone_demo} shows the configurations of the region queries, where the output of ring query is obtained by subtracting the outputs from D-ReZero model with the upper and lower bounds of the distance window. We can observe that ReZero is able to handle all types of query regions and different numbers of target speakers within the query regions, proving its effectiveness and potential for the general R-SE task.

\begin{table}[!t]
  \caption{Configurations of different region queries of the conical region query example. } 
  \label{tab:cone_demo}
  \centering
  \small
  \begin{tabular}{c|c|l|l|c}
    \toprule
    \textbf{ID} &
    \textbf{Query type} &
    \textbf{Query} $\mathbf{r}_a$ &
    \textbf{Query} $\mathbf{r}_s$ &
    \textbf{Q} \\
    \midrule
    1 & angular & $[-270\degree,-110\degree]$ & - & 2\\
    2 & spherical & -- & $[0,0.5]$ & 1 \\
    3 & ring & -- & $[0,1.1]$ - $[0,0.5]$ & 1 \\
    4 & conical & $[-150\degree,-90\degree]$ &$[0,1.5]$ & 1 \\    
    5 & conical & $[90\degree,120\degree]$ & $[0,1.0]$ & 1\\
    6 & conical & $[-130\degree,-60\degree]$ &$[0,0.6]$ & 0\\    
    7 & conical & $[-30\degree,30\degree]$ & $[0,1.0]$ & 0\\
    \bottomrule
  \end{tabular}
\end{table}

\section{Conclusion}
\label{sec:conclusion}



In this paper, we proposed \textit{\textbf{Re}}gion-customi\textit{\textbf{Z}}able sound \textit{\textbf{e}}xt\textit{\textbf{r}}acti\textit{\textbf{o}}n (\textit{\textbf{ReZero}}), a general and flexible framework for the region-wise sound extraction (R-SE) task. ReZero made use of spectral, spatial and region features to extract all target sources within a user-defined query region, which can be angular (angle window), spherical (distance threshold), or conical (angle window with distance threshold). A modified band-split RNN (BSRNN) model was also proposed for improved modeling of such features. Comprehensive experiment results showed that ReZero was able to properly handle different types of microphone array geometries, region query types, and performed consistently better than other benchmarking systems.

There are several future works to investigate. First, we only tested the performance of ReZero on small-scale microphone arrays, and its performance on large-scale arrays or ad-hoc arrays needs further validation. Second, how to properly define region features with large-scale arrays, especially when the target sources can locate inside the microphone array (e.g., in-car scenarios), is also important. Third, here we only considered region queries with regular shapes, and how to extend ReZero to support irregular-shaped region queries is also an interesting topic to study.


\bibliographystyle{IEEEbib}
\bibliography{refs}

\end{document}